\documentclass[9pt,conference,letter]{IEEEtran}
\IEEEoverridecommandlockouts

\usepackage{amsfonts}   
\usepackage{amssymb}  
\usepackage{amsmath}  
\usepackage{amsthm}
\usepackage{color}
\usepackage{graphicx}
\usepackage[numbers,sort&compress]{natbib}
\def\BibTeX{{\rm B\kern-.05em{\sc i\kern-.025em b}\kern-.08em
T\kern-.1667em\lower.7ex\hbox{E}\kern-.125emX}}
\usepackage{enumitem}
\usepackage{soul}
\usepackage{stmaryrd}
\usepackage{bm}
\usepackage{multirow}
\usepackage[aboveskip=2pt]{subcaption}
\usepackage{lipsum}
\usepackage{hyperref}
\usepackage{algorithm}
\usepackage{algpseudocode}
\usepackage{tabularx}
\usepackage{diagbox}

\newcommand \revis[1] {{\color{junglegreen} #1}}

\long\def\revis#1{\textcolor{black}{#1}}

\input{mysymbol.sty}

\title{
Learnable Digital Twin for Efficient Wireless Network Evaluation
    \thanks{%
        This research was sponsored by the Army Research Office and was accomplished under Cooperative Agreement Number W911NF-19-2-0269. 
        The views and conclusions contained in this document are those of the authors and should not be interpreted as representing the official policies, either expressed or implied, of the Army Research Office or the U.S. Government. 
        The U.S. Government is authorized to reproduce and distribute reprints for Government purposes notwithstanding any copyright notation herein.\\
        E-mails: \{boning.li, tde1, ak109, segarra\}@rice.edu, \{gunjan.verma.civ, ananthram.swami.civ\}@army.mil, jose.cortes2@mavs.uta.edu.
    }
}

\author{\IEEEauthorblockN{
Boning Li\IEEEauthorrefmark{1},
Timofey Efimov\IEEEauthorrefmark{1}, 
Abhishek Kumar\IEEEauthorrefmark{1},
Jose Cortes\IEEEauthorrefmark{2},
Gunjan Verma\IEEEauthorrefmark{3}, 
Ananthram Swami\IEEEauthorrefmark{3}, and
Santiago Segarra\IEEEauthorrefmark{1}}
\IEEEauthorblockA{
        \IEEEauthorrefmark{1}Rice University, USA
        \hspace{1em}
        \IEEEauthorrefmark{2}University of Texas at Arlington, USA
        \hspace{1em}
        \IEEEauthorrefmark{3}US Army DEVCOM Army Research Lab., USA\\
    }
    \\[-3.0ex]
}

\begin{document}
\setlength{\abovedisplayskip}{3pt}
\setlength{\belowdisplayskip}{3pt}

\maketitle

\begin{abstract}
Network digital twins (NDTs) facilitate the estimation of key performance indicators (KPIs) before physically implementing a network, thereby enabling efficient optimization of the network configuration. 
In this paper, we propose a learning-based NDT for network simulators.
The proposed method offers a holistic representation of information flow in a wireless network by integrating node, edge, and path embeddings. 
Through this approach, the model is trained to map the network configuration to KPIs in a single forward pass.
Hence, it offers a more efficient alternative to traditional simulation-based methods, thus allowing for rapid experimentation and optimization. 
Our proposed method has been extensively tested through comprehensive experimentation in various scenarios, including wired and wireless networks. 
Results show that it outperforms baseline learning models in terms of accuracy and robustness. 
Moreover, our approach achieves comparable performance to simulators but with significantly higher computational efficiency.

\end{abstract}


\section{Introduction}\label{s:intro}

Over the last two decades, digital twin technology has emerged as a highly effective approach for virtually modeling the physical world.
Offering rapid results, reduced resource consumption for proof-of-concept validation, and high accuracy, digital twins have gained significant traction in various domains, such as manufacturing, construction, and operations~\cite{madni2019leveraging,jones2020characterising,tao2018digital}. 
\revis{In particular, network digital twins (NDTs) have become popular in recent years,
driven by the demands of increasingly complex networks due to scale and heterogeneity, as well as the growing availability of data from sensors, simulators, and other sources~\cite{wang2021unmanned,zhao2022elite}.
NDTs have enormous untapped potential in wireless networks and are of particular interest in military settings, where large, dynamic, wireless networks may exhibit behaviors that are difficult to predict and model, and hence optimize, a priori.}
For a wide range of downstream tasks like network maintenance and management, NDTs usually play the role of predicting key performance indicators (KPIs) given the network topology, routing scheme, traffic data, and other relevant features~\cite{rusek2020routenet,groshev2021toward,baert2021digital}.\looseness=-1

NDTs can be implemented using neural networks trained with data collected from the actual or simulated communication processes~\cite{khan2022digital}.
Upon the completion of training, the NDT model can be deployed to monitor the traffic in the network and adjust routing in real time around congested areas of the network~\cite{kumar2018novel,lin2021stochastic,poularakis2021generalizable} or to prioritize certain types of traffic~\cite{bellavista2021application}.
This can be especially useful in dynamic environments where traffic patterns are constantly changing. 
Compared to  traditional network simulators such as ns-3~\cite{henderson2008network} or OMNeT++~\cite{varga2010overview}, pretrained models have vastly improved run-time efficiency, requiring a forward pass on the order of milliseconds instead of the costly packet-level simulations that can take on the order of minutes.
However, while topology-agnostic learning models may be plausible for wired networks~\cite{sun2019gradientflow,tariq2013answering}, they are definitely not suitable for wireless settings due to the time-varying nature of the wireless topology. 
In many wireless networks, such as those used in military-relevant settings, devices can move, and the channels change over time, making it essential to handle different network topologies dynamically.\looseness=-1

Graph neural networks (GNNs) and their variants, given their ability to process topological information, have rapidly emerged as a favored machine learning tool for communication network applications~\cite{geyer2019deepcomnet,xiao2018deepq,wang2021neuralmon}. 
For instance, \cite{geyer2019deepcomnet} proposed a GNN framework to predict the average delay and throughput of TCP flows based only on a graph-based representation of network topologies.
The work in~\cite{poularakis2021generalizable} used the graph attention (GAT) model to predict network congestion but did not investigate its generalizability on different topologies. 
Routenet, introduced in \cite{rusek2020routenet} \revis{as a seminal work in the space of learning-based NDTs,} proposes a neural architecture with learnable link and path embeddings for network KPI prediction. 
Indeed,~\cite{rusek2020routenet} estimated network delay, jitter, and packet loss to optimize routing and update the network under budget constraints.
Additionally,~\cite{happ2021graph} estimated network delay for heterogeneous scheduling policies, and~\cite{ferriol2022routenet} and~\cite{ferriol2022routenetfermi} extended RouteNet to take queueing information for more complex traffic and network modeling.
\revis{However, all of these works were evaluated only in wired settings and the proposed models lack the ability to capture wireless phenomena such as interference.}
These limitations have prevented RouteNet and its extensions from achieving good performance in the context of wireless networks.\looseness=-1

In this regard, we present PLAN-Net, a graph-based 
neural architecture with learnable \underline{P}ath, \underline{L}ink, \underline{A}nd \underline{N}ode embeddings. 
Our novel approach overcomes existing methods' reliance on wired topologies while still harnessing the concept of RouteNet's utilization of link-path information. 
By incorporating explicit graph learning techniques with edge-conditioned graph convolutional network (GCN) layers, we are able to effectively model wireless network performance.
Another key advantage of PLAN-Net is its ability to dramatically enhance computing efficiency when compared to simulator-based NDTs. 
This improvement in efficiency is particularly important, as it enables faster and more effective decision-making in complex systems. 
This makes PLAN-Net an ideal choice for mission-critical applications where speed, accuracy, and efficiency are critical factors for success.

\section{Problem Formulation}\label{s:pre}

\begin{figure}[t]
\centering
    \includegraphics[width=\linewidth,trim= 0cm 16.5cm 0 0cm]{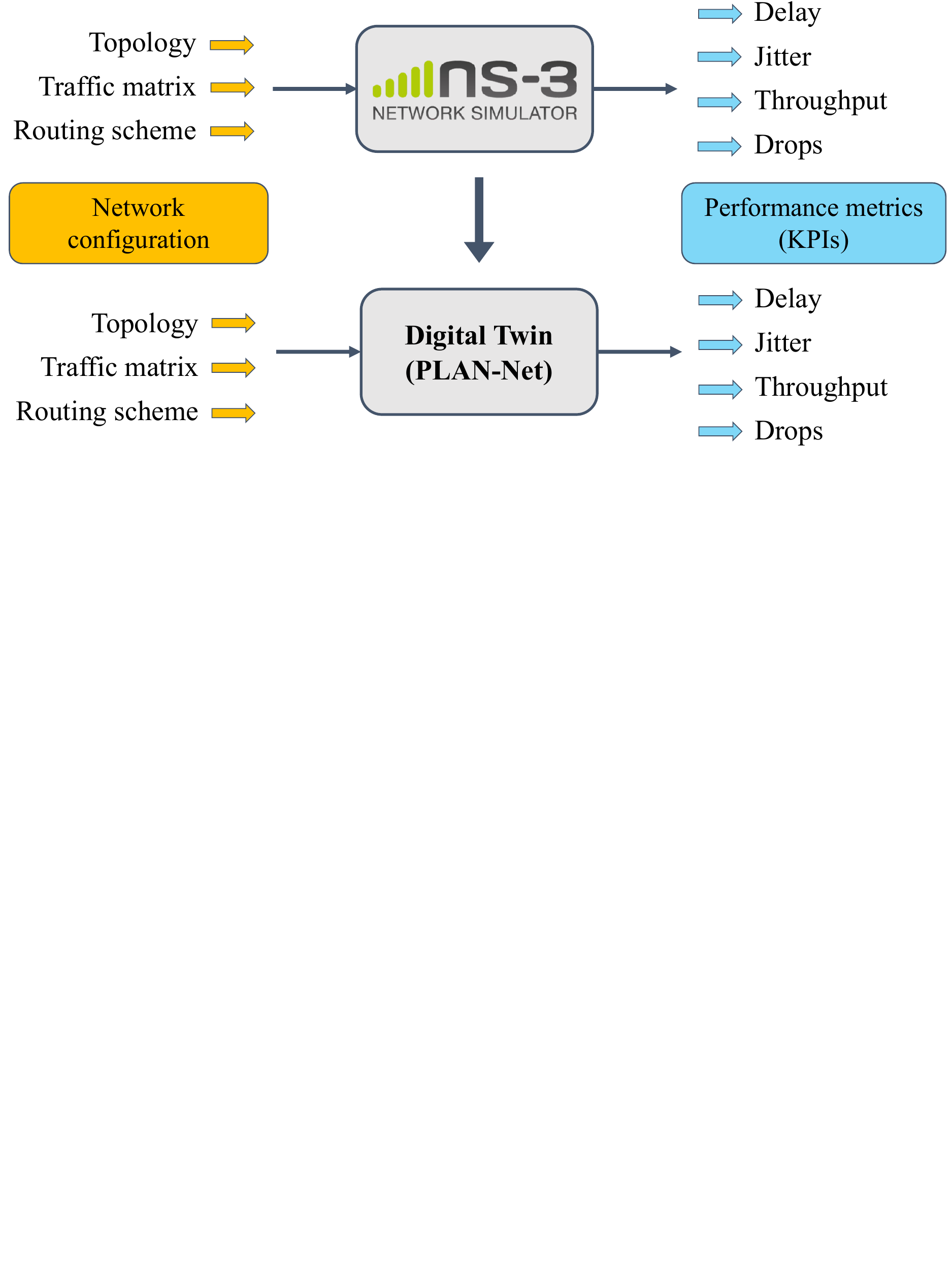}
    \caption{
    A network simulator can be replaced with a more efficient learning-based NDT in predicting KPIs for a given network.
    }
    \label{f:overview}
\end{figure}

While network simulators can aid the design of real-world systems, one can also use a learning-based NDT for higher efficiency by avoiding running the entire simulation pipeline for every unseen network instance.
As the schematic in Fig.~\ref{f:overview} illustrates, a  network simulator takes as input the network topology with a set of network protocols, a traffic matrix, and a source-destination routing scheme.
As output, for every path characterized by a source-destination pair, it estimates several KPIs by monitoring those statistics from simulating activities of interest for some duration.
The same set of input and output applies to our proposed NDT model.
However, instead of relying on simulation, PLAN-Net estimates the KPIs through a specialized neural architecture, which can be \revis{3 to 4 orders of magnitude} 
more efficient than a simulator in practice.
The rest of this section gives more details about the system input and output.

Let us denote the \textit{topology} of an ad hoc network as a graph $\ccalG{\,=\,}(\ccalN,\ccalL)$, where $\ccalN$ represents the set of nodes and $\ccalL$ the set of links.
For easier interpretation of the idea, let us assume that $\ccalG$ is connected and set aside considerations of interference for now.
Nodes communicate with each other through links, which form a path from a source node to a destination node, either as a single link or a sequence of links.
Data are generated by applications on source nodes (with possibly different rates) and transmitted to destination nodes, creating \textit{traffic} flows in the network. 
This process can be characterized by a tuple $\bbt{\,=\,}(t_\text{on}, t_\text{off})$, indicating that the node generates traffic for a duration of $t_\text{on}$ seconds, pauses for $t_\text{off}$ seconds, and switches between this on-and-off pattern throughout the simulation.
In order to mimic the indeterminacy in real-life user behaviors, we sample these on-off times from exponential distributions. 
The traffic matrix can be formally written as $\bbT{\,=\,}[\bbtau_p]_{p=0}^{P-1}{\,\in\,}\mbR^{P\times2}$ with $\bbtau$ representing the mean values of the exponential distributions from which $\bbt$ are sampled, and $p$ indexing $P$ source-destination pairs (further details in Section~\ref{s:exp}).
To determine which links the data packets should traverse, we need a \textit{routing} scheme, such as the optimized link state routing (OLSR), a commonly used protocol based on the shortest-path algorithm.
While the simulator can discover paths according to any routing protocol that has been programmed, the NDT lacks this ability; therefore, we give the simulator-determined set of paths, denoted as $\ccalP$, as a part of the input to PLAN-Net.

In terms of the output, we predict delay, jitter, throughput, and packet drops, which can comprehensively evaluate the network performance.
In this context, \textit{delay} (in milliseconds) \revis{refers to the end-to-end delay per packet, calculated for each flow as the sum of all delays for received packets divided by the number of received packets}. 
It effectively relates to the responsiveness of various applications.
\textit{Jitter}, or delay variation, is defined as the difference between the delays of two successive packets.
It measures fluctuation in the delay of packets as they traverse the network, which is critical for real-time applications, such as \revis{battlefield communications and unmanned aerial vehicles.}
\textit{Throughput} (in $\text{kb/s}$) refers to the amount of data transmitted or received within a given time period, directly impacting the efficiency and capacity of a network.
Higher throughput is generally preferred for, e.g., file transfer, cloud computing, \revis{and streaming video for search-and-rescue operations}. 
Finally, the count of discarded or lost data packets during transmission, namely \textit{packet drops}, indicates the health of the network.
Drops can occur due to various causes and can significantly impact network reliability. 

In summary, easy, fast, and accurate acquisition of network KPIs is crucial for network administrators to proactively detect network congestion, identify faulty components, and optimize network resources, all of which contribute to maintaining optimal network performance and delivering a satisfactory user experience.
Training on the input and output described above, we develop an NDT with specialized neural architecture to learn
the desired mapping.

\section{Proposed Method}\label{s:alg}

\begin{figure}[t]
\centering
    \includegraphics[width=\linewidth,trim= 0 0 .3cm 7cm]{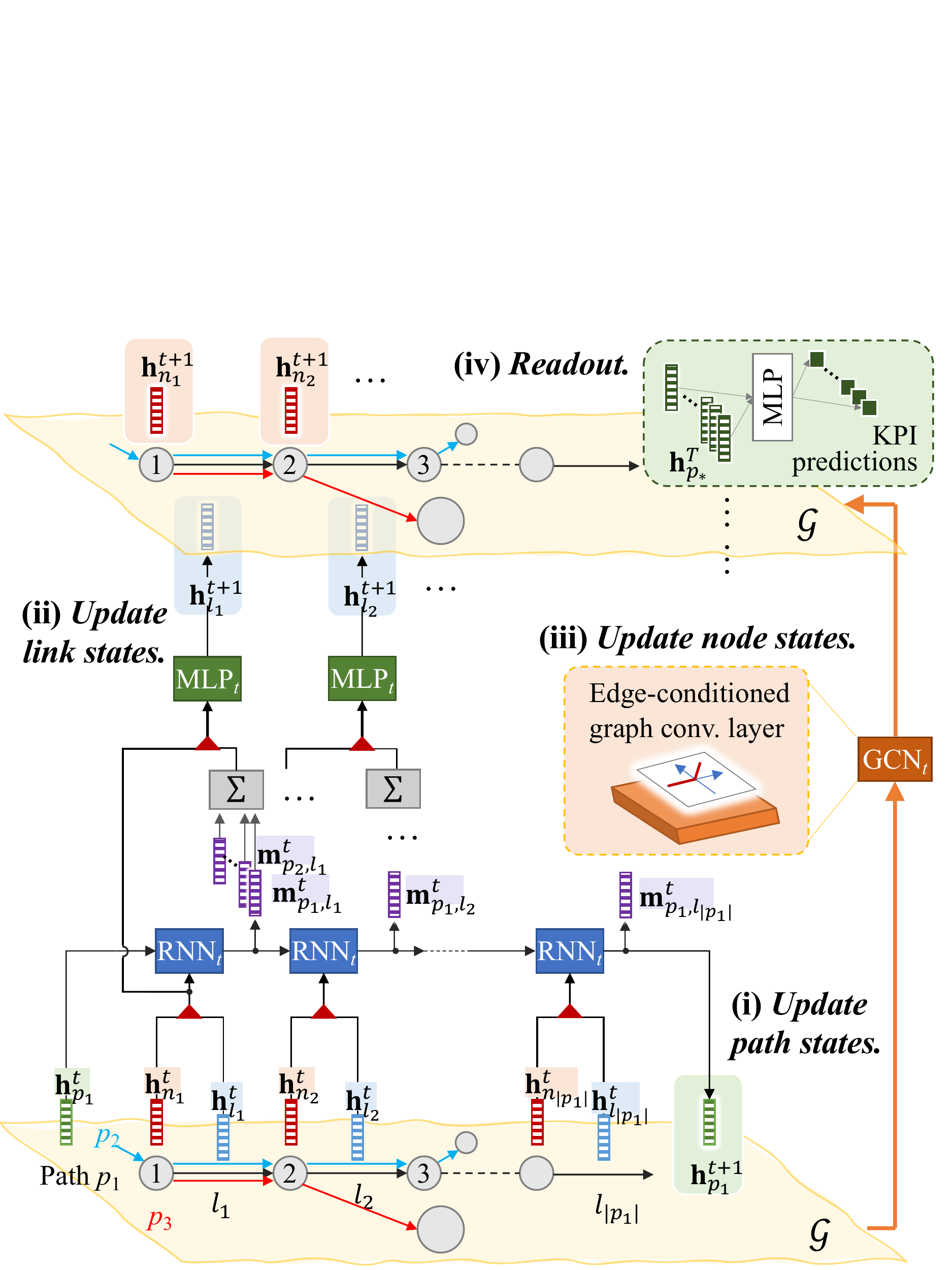}
    \caption{
    Architecture of the proposed PLAN-Net. 
    }
    \label{f:scheme}
\end{figure}

\begin{figure}[t]
\centering
    \includegraphics[width=.75\linewidth, trim=0cm 13.5cm 3cm .5cm]{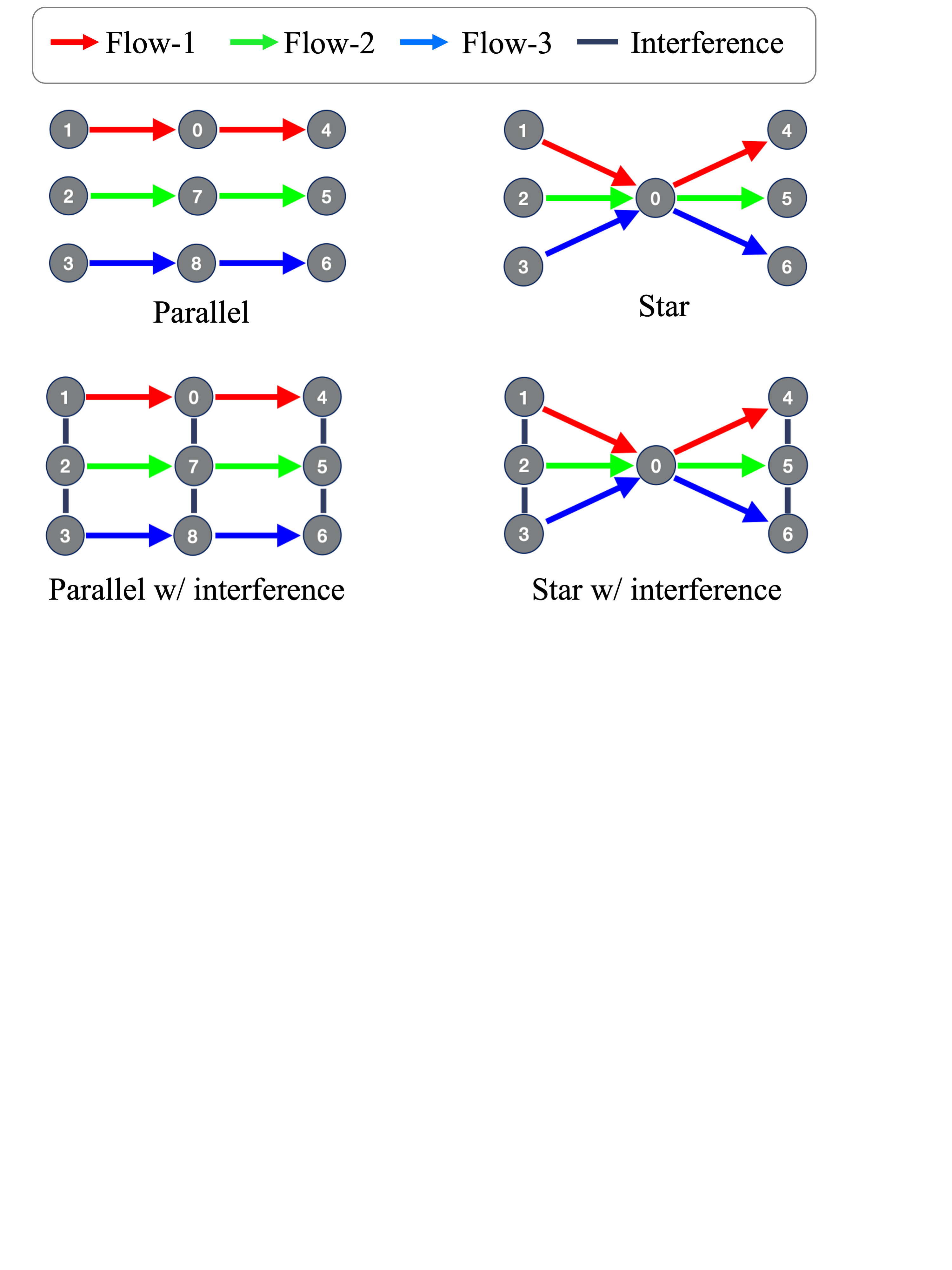}
    \caption{
    Parallel versus star topologies, without and with interference.
    }
    \label{f:para-star}
\end{figure}

In this paper, we propose PLAN-Net, a graph-based neural network model for predicting KPIs in \revis{wired or wireless} communication networks.
As illustrated in Fig.~\ref{f:scheme}, the proposed architecture unifies path, link, and node embeddings to effectively capture the interrelationships among them. 
In particular, our approach enhances the existing link-path hybrid RouteNet model~\cite{rusek2020routenet} by incorporating additional node embeddings through edge-conditioned GCNs, which can improve performance on the four scenarios shown in Fig.~\ref{f:para-star}.
For example, RouteNet treats parallel and star topologies as equivalent and overlooks interference between different flows. 
Essentially, since none of the three flows share a link in any of the four configurations, an architecture that only considers link and path embeddings cannot distinguish between these configurations.
In contrast, our \revis{GCN-based node embeddings} enable the model to consider {\it inter}-path topological information such as conflicts and interference, rather than relying solely on the aggregated {\it intra}-path link information. 

Referring back to Fig.~\ref{f:scheme}, in each PLAN-Net layer, network embeddings are updated in a three-step manner.
First, the path embedding is updated through an RNN, which is contingent upon the preceding link and node embeddings.
The choice of RNN is dictated by its ability to capture dependence in sequences of variable length.
As the length of paths within the wireless network may vary, RNN is well-suited to handle this variability.
Second, the intermediate outputs of the previously mentioned RNN at each link are fed into an MLP, along with the prior link and node embeddings.
This step updates link embeddings and, by employing the MLP, our model can effectively combine information from multiple sources with reduced dimensionality. 
Finally, the node and link embeddings are aggregated by an edge-conditioned graph convolutional layer for new node embeddings, thus incorporating the graph topology into the process of estimating embeddings. 
It first aggregates the link embeddings that are going out of the node and then performs an update with the concatenation of the node and aggregated link embeddings. 
In summary, this iterative approach allows the model to capture complex dependencies within the wireless network, leading to its ability to learn a comprehensive representation of the graph structure, which in turn improves the accuracy of KPI predictions.
Thus, we have effectively addressed the limitation of RouteNet by introducing node embeddings that influence the updates for link and path embeddings. 

Algorithm~\ref{alg:cap} outlines the step-by-step procedure that PLAN-Net follows. 
Our algorithm takes as input the network graph $\ccalG$, which comprises a node set $\ccalN$ and a link set $\ccalL$, along with a list of paths $\ccalP$. 
When there is a link between two nodes, it signifies that a direct communication link has been established between them.
Each path $p$ is defined by \revis{an ordered sequence} of links that starts from the source node and ends at the destination node. 
In this sense, the statement $l{\,\in\,}p$ is true if $l$ is in that sequence.
Informally speaking, we may say a node $n$ is in path $p$, or $n{\,\in\,}p$, if any $l{\,\in\,}p$ transverses $n$.
Initial embedding states are also required for every path, link, and node entry in the network, incorporating prior knowledge such as the traffic vector $\bbtau_p$, link capacity $c_l$, and node degrees $d_n$, respectively.
The initial embeddings are vectors with the first element as the aforementioned prior values and the rest zero-padded.
The algorithm proceeds iteratively for $T$ iterations, with each iteration consisting of the three-step updates discussed earlier and illustrated in Fig.~\ref{f:scheme}.
After completing the iterations, the algorithm applies \revis{several MLP layers} to the final path embedding $\mathbf{h}_p^T$ to {\it readout} the output $y_p$, which represents the estimated KPI for one of the four aforementioned metrics (delay, jitter, throughput, or drops), for path $p{\,\in\,}\ccalP$ in the wireless network.

\begin{algorithm}
\caption{PLAN-Net algorithm.}\label{alg:cap}
\begin{algorithmic}[1]
\Require Graph $\ccalG=(\ccalN, \ccalL)$, paths list $\ccalP$
\Ensure $\bbtau_p, c_l, d_n > 0$
\State $\bbh_p^0{\,\gets\,}[\bbtau_{p}, 0, \cdots, 0]^\top,\,\forall\, p\in\ccalP$
\State $\bbh_l^0{\,\gets\,}[c_l, 0, \cdots, 0]^\top,\,\forall\, l\in\ccalL$
\State $\bbh_n^0{\,\gets\,}[d_n, 0, \cdots, 0]^\top,\,\forall\, n\in\ccalN$
\For{$t= 0, 1, \cdots, T-1$}
\State \textbf{(i)~\textit{Update path states.}}
\For{every path $p$ in $\ccalP$}
    \For{every ordered link $l$ in $p$}
        \State $\bbh_p^t{\,\gets\,}\RNN_t (\bbh_p^t, \cat[\bbh_l^t, \bbh_n^t])$, where $n{\,=\,}\src(l)$
        \State\hfill\Comment{$n$ is the source node of $l$}
        \State $\bbm_{p,l}^{t}{\,\gets\,}\bbh_p^{t}$
    \EndFor
    \State $\bbh_p^{t+1}{\,\gets\,}\bbh_p^{t}$
\EndFor
\State \textbf{(ii)~\textit{Update link states.}}
\For{every link $l$ in $\ccalL$}
    \State $\bbh_l^{t+1}{\,\gets\,}\MLP_t (\cat[\bbh_l^t, \bbh_n^t, \agg_{p}\{\bbm_{p,l}^t{\,|\,}l{\,\in\,}p\}])$
    \State\Comment{$p$ is all paths that contain $l$}
\EndFor
\State \textbf{(iii)~\textit{Update node states.}}
\For{every node $n$ in $\ccalG$}
    \State $\bbh_n^{t+1}{\,\gets\,}\GCN_t (\cat[\bbh_n^t, \agg_l\{\bbh_l^t{\,|\,}l{\,=\,}L^{+}(n)\}]; \ccalG)$ 
    \State\hfill\Comment{$l$ is all links out of $n$}
\EndFor
\EndFor
\State \textbf{(iv)~\textit{Readout.}}
\State $y_p = \MLP(\bbh_p^{T})$
\end{algorithmic}
\end{algorithm}

The main motivation for incorporating graph-based learning for node embeddings is to achieve higher generalizability across topologies.
This is necessary for better representation of densely connected and highly congested nodes, which was not previously accounted for \revis{by RouteNet.} 
By leveraging GCNs to learn node embeddings, our approach provides a more comprehensive representation of \revis{the wireless traffic conflicts and node interference,} leading to improved prediction accuracy.

\section{Numerical experiments}\label{s:exp}

Our experiments\footnote{The code is available at \texttt{ https://github.com/bl166/wireless\\\_digital\_twin\_milcom}.} consist of three parts, each featuring different network settings and thorough comparisons between PLAN-Net and other baselines. 
In Section~\ref{ss:wired}, we focus on a basic wired network under different levels of congestion. 
In Section~\ref{ss:wifi}, we include wireless channels and more complex topologies of various densities. 
Section~\ref{ss:wifi-pert} further explores the generalizability of PLAN-Net across perturbations in the network topology. 

\subsection{Wired communication network}\label{ss:wired}

\begin{figure}[t]
    \centering
    \begin{subfigure}[t]{.32\linewidth}
        \centering
        \includegraphics[width=\linewidth, trim=.8cm 1.2cm .9cm 1.2cm, clip]{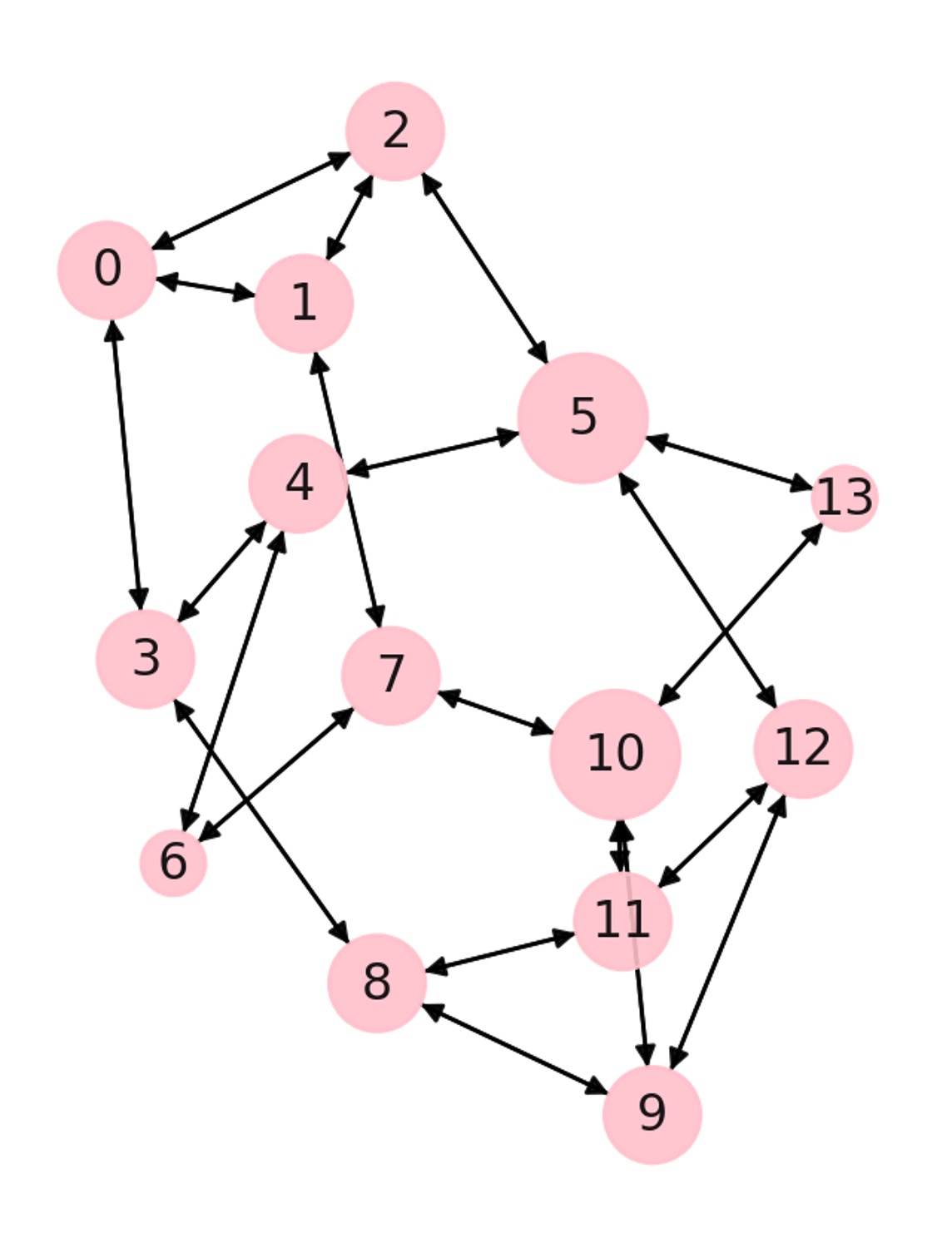}
        \caption{Links.}\label{ff:11}
    \end{subfigure}%
    ~\hfill%
    \begin{subfigure}[t]{.32\linewidth}
        \centering
        \includegraphics[width=\linewidth, trim=.8cm 1.2cm .9cm 1.2cm, clip]{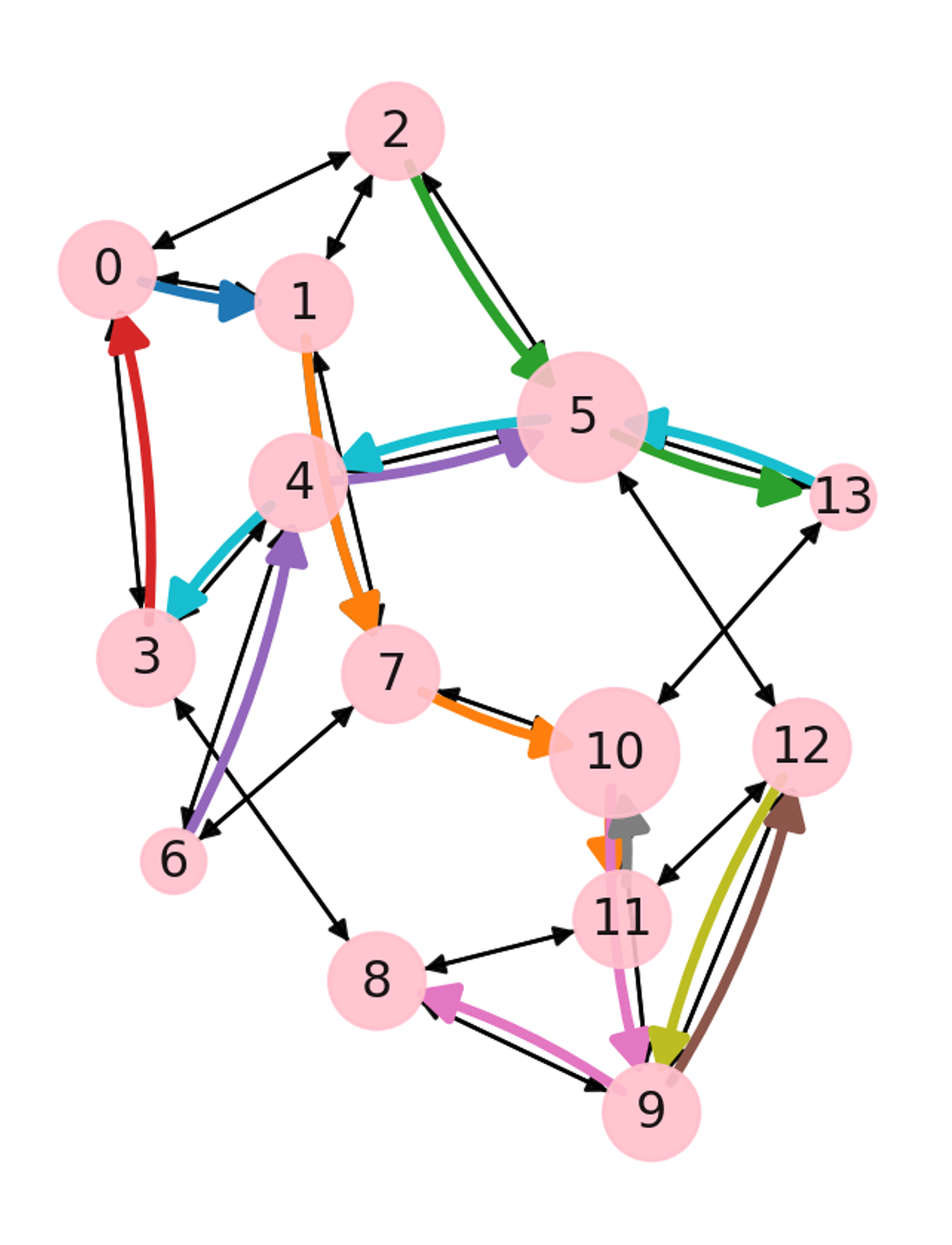}
        \caption{Paths.}\label{ff:12}
    \end{subfigure}%
    ~\hfill%
    \begin{subfigure}[t]{.32\linewidth}
        \centering
        \includegraphics[width=\linewidth, trim=.8cm 1.2cm .9cm 1.2cm, clip]{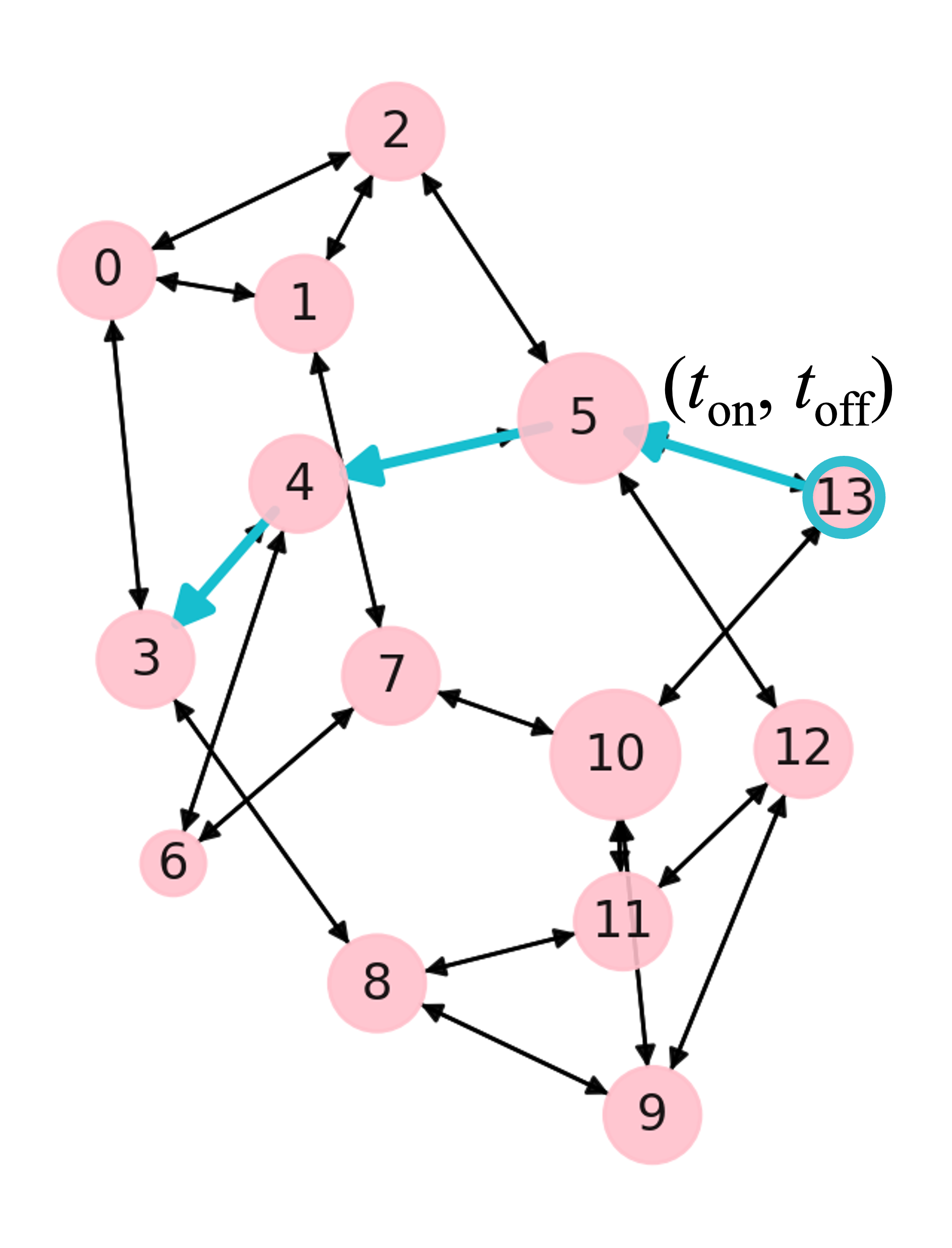}
        \caption{Traffic.}\label{ff:13}
    \end{subfigure}%
    \caption{
    NSFNet-based data overview.
    }
    \vspace{-1.5em}
\end{figure}

To start with, we introduce the NSFNet~\cite{Claffy1993Traffic}, a commonly used benchmark topology in wired network research and simulation studies~\cite{Chen2020Routing,rusek2020routenet}.
Its topology (Fig.~\ref{ff:11}) contains $|\ccalN|{\,=\,}14$ nodes with $|\ccalL|{\,=\,}42$ fixed and unweighted directed links.
We manually select 10 distinct pairs of nodes as the source and destination of $|\ccalP|{\,=\,}10$ paths in Fig.~\ref{ff:12}.
A path is acquired by the OLSR protocol given the underlying topology, defined as a sequence of links carrying traffic flow characterized by $(t_\text{on},t_\text{off})$ at the source node of that path (see Fig.~\ref{ff:13}). 
We compute paths assuming that no other data traffic is traversing the network and fix the OLSR routing table to simplify the problem.
The path lengths range from 1 to 3 hops.
We begin by studying the wired setting.

In terms of the traffic data, at the source node of each path, we sample the on/off times $t_\text{on}$ and $t_\text{off}$ from exponential distributions $\text{Exp}(1/\tau_\text{on})$ and $\text{Exp}(1/\tau_\text{off})$, respectively, whose means $\tau_\text{on}$ and $\tau_\text{off}$ are sampled from the set $\mbT{\,=\,}\{1,10,20\}$ with uniform probabilities.
During the `on' phases which repeatedly last for $t_\text{on}$ and pause for $t_\text{off}$ seconds, data are generated and transmitted from the source nodes to the \revis{destination} 
nodes at a predefined data rate in the set $\{50, 75, 100, 125, 150\}\text{ kb/s}$.
During the simulation, ns-3 monitors several performance metrics for each path, including the number of transmitted and received packets, which can be used to compute packet drops, as well as delay, jitter, and throughput. 
These metrics serve as {\it ground truth} labels for our prediction model.
The {\it input} data, relevant to the traffic in the network, comprises the network topology $\ccalG$ and the mean on/off times associated with all $p{\,\in\,}\ccalP$, represented as a matrix $\bbT{\,\in\,}\mbT^{P{\times}2}$ with $P{\,=\,}|\ccalP|$.
For each choice of data rate, we obtain 1500 training samples, which are further split for 3-fold cross-validation (CV), and 1000 test samples on which we evaluate the predictions.
On the holdout test set, the predicted KPI for a test sample is obtained by averaging the predictions of the top-performing models from all CV folds.\looseness=-1

Recalling the details of our architecture (Fig.~\ref{f:scheme}), the implemented PLAN-Net is comprised of $T{\,=\,}3$ layers, through which it learns path, link, and node embeddings of different dimensionalities. 
Specifically, path RNNs learn 32-dimensional path embeddings, link MLPs learn 16-dimensional link embeddings via four hidden layers of sizes $\{32, 64, 128, 32\}$, and node GCNs aggregate information from one-hop neighbors to update 16-dimensional node embeddings. 
Finally, the readout MLP uses three hidden layers of sizes $\{64, 32, 16\}$ to project the path embeddings into one-dimensional output, followed by a linear activation.
We use the Adam optimizer to optimize a mean squared error (MSE) loss function with L2 regularization, whose hyperparameters are determined and adjusted \revis{separately for each performance metric.} 
Separate models are trained from scratch to predict different KPI metrics.

In addition to PLAN-Net, we also investigate the following baseline methods in two broad categories, namely the learning-based NDTs, including RouteNet and the generic GNN models, as well as the simulation-based ns-3 methods.

\begin{enumerate}[wide,label={\arabic*)},labelindent=0pt]
    \item RouteNet. 
    It follows a similar architecture to PLAN-Net, but does not incorporate node embeddings or explicit graph learning components. 
    \revis{Its ability to capture link-path relational information is suitable 
    for} wired settings, where there is less interference and fewer conflicts.
    
    \item GNN. 
    It is a basic graph neural network that predicts a multi-dimensional output from the learned graph representation for all paths.
    The node features, denoted as $\bbx^{(j)}$ for node $n_j$, have $2P$ dimensions to incorporate traffic information of all paths in a fixed order:\looseness=-1 
    \begin{equation*}
        x^{(j)}_i=
        \begin{cases}
          \tau_{p_k\text{-on}}, & \text{if } n_j\in p_k, \text{ and } i=2k\\
          \tau_{p_k\text{-off}}, & \text{if } n_j\in p_k, \text{ and } i=2k+1\\
          0, & \text{otherwise, }\forall\,k=0,...,P{-}1.
        \end{cases}
    \end{equation*}
    However, it has a fatal defect that the output dimension is predefined, thus limiting its usage in practice due to \revis{variability in the number or length of paths}. 
    \item ns-3. 
    The simulator can provide a natural benchmark for evaluating KPI predictions.
    We can feed the same topology, traffic, and path data to ns-3 and let it simulate, once or multiple times again, what the KPIs may be.    
    \revis{Running multiple ns-3 simulations and averaging the predictions can mitigate the variability caused by the random sampling of $t_\text{on}$ and $t_\text{off}$ from the same distribution,
    thus increasing the accuracy of the model within the limits determined by the inherent uncertainties in the system being modeled.}
    In addition to the single-run results, we also report ns-3$^+$ and ns-3$^{+\!+}$, which represent the results obtained by averaging the single-run results with one and two additional runs, respectively. 
\end{enumerate}

Although our ground-truth KPIs are generated using ns-3, the inherent stochasticity of the simulation implies that calling ns-3 again with the same inputs will result in different KPIs.
Thus, \emph{the ns-3 methods will yield non-zero testing errors, which will be naturally reduced by averaging more simulations} (as in ns-3$^+$ and ns-3$^{+\!+}$).
Regarding the input traffic data, it should be pointed out that the input to both NDTs and ns-3 can be represented by a matrix of mean times $\tau$.
While internally in ns-3 we sample the instantaneous on/off times $t$ from $\text{Exp}(1/\tau)$ every time it runs a simulation, the sampled $t$ values are never revealed to NDTs. 

\begin{figure}[t]
    \centering
    \begin{subfigure}[t]{\linewidth}
        \centering
        \includegraphics[width=\linewidth, trim=.25cm .3cm .33cm .5cm]{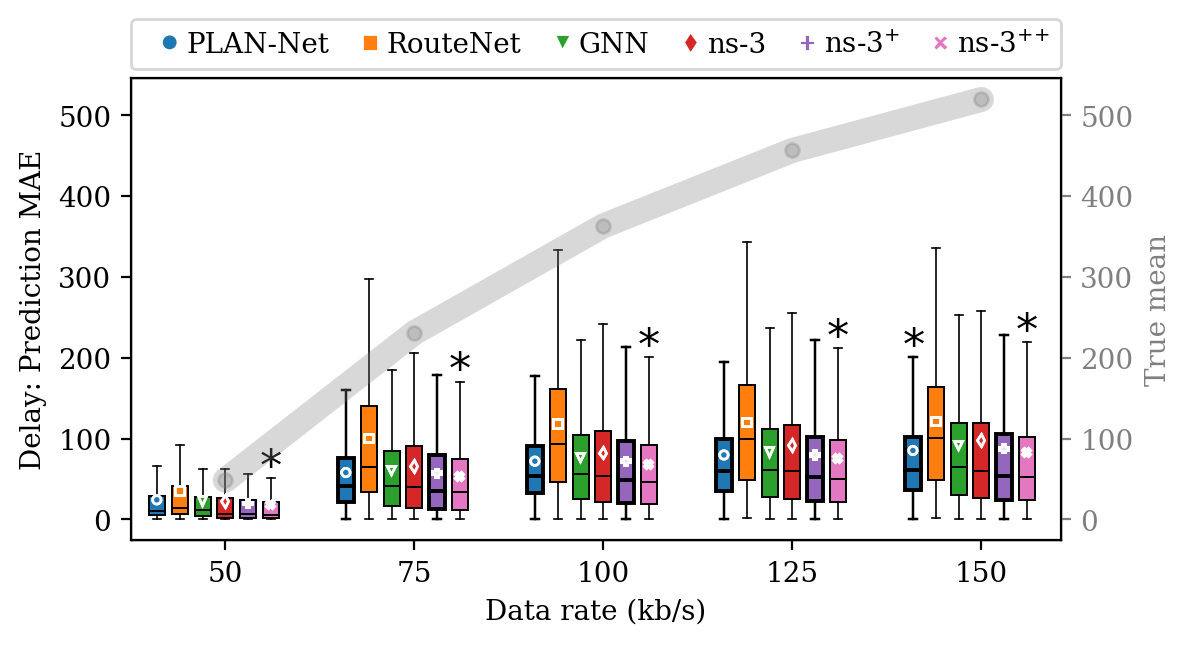}
    \end{subfigure}%
    \caption{
    Box plots of delay prediction MAE by candidate methods with mean delay values overlaid. 
    Asterisks placed on top of the boxes indicate the presence of statistical significance of the corresponding methods' mean MAE being lower than the others in each group.
    The curve represents the mean values of the ground truth delay for each data-rate group.
    }
    \label{ff:1dr}
    %\vspace{-1em}
\end{figure}

Fig.~\ref{ff:1dr} shows box plots of the mean absolute errors (MAE) of all candidate methods in predicting path delays. 
It is plotted against different data rates that reflect varying degrees of network congestion. 
The first notable observation is that PLAN-Net performs the best among all the learning models under all network conditions.
It surpasses not only generic models like GNN but also RouteNet, another specialized architecture for network evaluation.
When juxtaposed with the ns-3 benchmarks, PLAN-Net exhibits comparable performance to ns-3$^+$ in networks with $75\text{ kb/s}$ or higher data rates.
Especially when the congestion gets severe (with a data rate of $150\text{ kb/s}$, for example), the performance gap between ns-3$^{+\!+}$ and PLAN-Net becomes insignificant.
This suggests that, by being trained on multiple simulated settings, PLAN-Net's outputs tend to be closer to the expected KPIs, an effect akin to averaging several simulations as in ns-3$^{+\!+}$.

Table~\ref{tab:1} provides numerical performance results for all KPIs (delay, jitter, throughput, and drops), for a $100\text{ kb/s}$ data rate.
The metric is the mean and standard deviation (SD) of the MAE \revis{between the inter-quartile-range (IQR) normalized true and predicted KPI values,} or the NMAE\footnote{\revis{Formally stated, $\text{NMAE}(\bby_t,\bby_p){\,=\,}\text{MAE}(\bby_t,\bby_p)/\text{IQR}(\bby_t)$, where $\bby_t$ is the true KPI values and $\bby_p$ the predicted values, and $\text{IQR}(\bby_t){\,=\,}\text{Q3}(\bby_t)-\text{Q1}(\bby_t)$ is the inter-quantile range of $\bby_t$.}} for abbreviation. 
\revis{By applying the IQR normalization, all KPI metrics are brought to a uniform scale, while preserving the relative order of performance by each method. 
Additionally, IQR normalization is robust to the presence of rare yet influential outliers without distorting the overall distribution of the results.}
It is clear that PLAN-Net is comparable to the 2-run ns-3$^+$ for all KPIs, with the strongest performance corresponding to throughput. 
Generally, ns-3$^{+\!+}$ provides the lowest prediction errors, and additional averaging may reduce the error even further. 
However, due to its time-consuming nature, it is often impractical to run ns-3 multiple times.

\begin{table}[t]
    \caption{
    Performance of candidate methods for multiple KPIs measured by NMAE on the test set.
    The data rate for both the training and test samples is fixed at $100 \text{ kb/s}$.
    }\label{tab:1}
    \centering
    \small
    \resizebox{\linewidth}{!}{%
        \begin{tabular}{|l|cc|cc|cc|cc|}
        \hline
             \multirow{2}{*}{\diagbox[width=2cm]{Method}{Metric}} &  \multicolumn{2}{c}{Delay} &  \multicolumn{2}{c}{Jitter} &  \multicolumn{2}{c}{Throughput} &  \multicolumn{2}{c|}{Drops} \\
        \cline{2-9}
             & Mean  & SD & Mean  & SD & Mean  & SD & Mean  & SD \\
        \hline
PLAN-Net & 0.18  & 0.19 & 0.29  & 0.47 & \textbf{0.17}  & \textbf{0.18} & 0.25  & 0.31 \\
RouteNet & 0.29  & 0.27 & 0.38  & 0.52 & 0.21  & 0.21 & 0.34  & 0.41 \\
GNN & 0.19  & 0.22 & 0.36  & 0.50 & 0.25  & 0.21 & 0.26  & 0.32 \\
ns-3 & 0.20  & 0.26 & 0.31  & 0.69 & 0.22  & 0.24 & 0.29  & 0.42 \\
ns-3$^{+}$ & 0.18  & 0.23 & 0.27  & 0.55 & 0.19  & 0.21 & 0.25  & 0.35 \\
ns-3$^{+\!+}$ & \textbf{0.17}  & \textbf{0.22} & \textbf{0.26}  & \textbf{0.51} & 0.18  & 0.20 & \textbf{0.24}  & \textbf{0.34} \\
        \hline
        \end{tabular}
    }
\end{table}

\subsection{Wireless network in regular grid topology}\label{ss:wifi}

\begin{figure}[t]
    \centering
    \begin{subfigure}[t]{.3\linewidth}
        \centering
        \includegraphics[width=\linewidth, trim=1.2cm 0 1.2cm 1cm, clip]{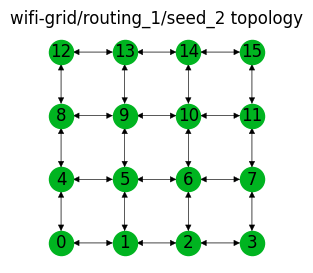}
        \caption{$P_\text{tx}{\,=\,}12$.}\label{ff:21}
    \end{subfigure}%
    ~\hfill%
    \begin{subfigure}[t]{.3\linewidth}
        \centering
        \includegraphics[width=\linewidth, trim=1.2cm 0 1.2cm 1cm, clip]{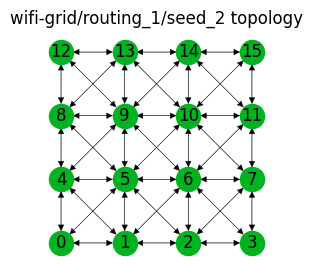}
        \caption{$P_\text{tx}{\,=\,}16$ (default).}\label{ff:22}
    \end{subfigure}%
    ~\hfill%
    \begin{subfigure}[t]{.3\linewidth}
        \centering
        \includegraphics[width=\linewidth, trim=1.2cm 0 1.2cm 1cm, clip]{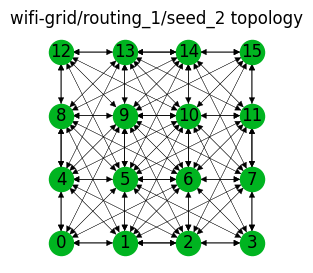}
        \caption{$P_\text{tx}{\,=\,}20$.}\label{ff:23}
    \end{subfigure}%
    \caption{
        Regular grid Wi-Fi network topology under different levels of transmit power $P_\text{tx}$ in dBm results in varying topological densities.
    }\label{ff:2}
\end{figure}

In the preceding experiment, we utilized prefixed links as opposed to Wi-Fi links. 
In this experiment, however, we will utilize Wi-Fi links based on the 802.11a model~\cite{yans-wifi}, albeit in an artificial use case. 
As shown in Fig.~\ref{ff:2}, 16 nodes are placed in a $4{\times}4$ square grid with $30$ meters between neighboring nodes on a row or column. 
The channel loss and delay models depend on the log distance propagation loss and the constant speed propagation delay, respectively.
From Fig.~\ref{ff:21} through~\ref{ff:23}, the density of this topology relies on the transmit power $P_\text{tx}$, which, if set higher, can result in more reliable links and shorter paths but also potentially stronger interference.
The figures presented in this context can also be interpreted as the adjacent nodes that are registered in the OLSR routing tables.
To clarify further, we model these topologies as weighted graphs \mbox{$\ccalG{\,=\,}(\ccalN, \ccalL, \ccalE)$}, where $e_{ij}{\,\in\,}\ccalE$ represents the weight associated with the link connecting nodes $n_i$ and $n_j$ such that 
\mbox{$e_{ij} = \log^{-1}[1+d(n_i, n_j)]$}, roughly characterizing the strength of the link between two nodes as a function of their distance. 
It is worth noting that interference and conflicts are crucial factors that can affect network performance, yet they are not explicitly modeled. 
Nonetheless, our research demonstrates that these factors can be learned implicitly through the introduction of node embeddings.

\begin{figure}[t]
    \centering
    \includegraphics[width=\linewidth, trim=.25cm 0.25cm 0.33cm 0.25cm, clip]{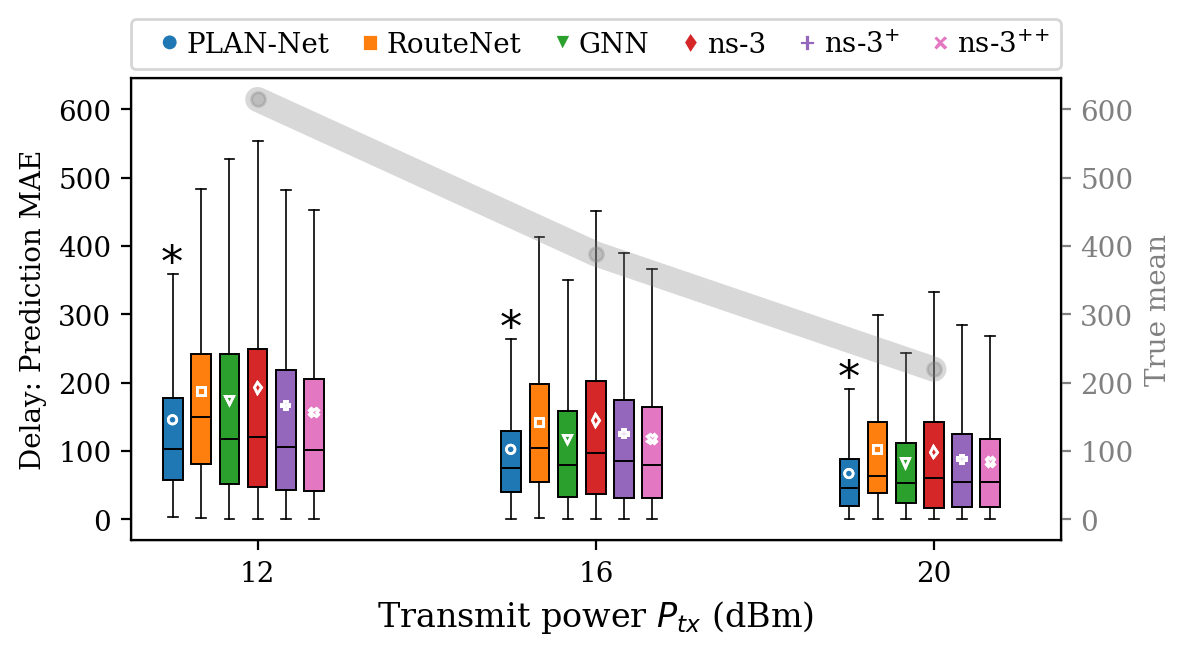}
    \caption{
    Box plots illustrating the variability in delay prediction MAE achieved using different methods in the grid Wi-Fi network, as measured by MAE statistics.
    The curve marks the mean values of the ground truth delay of each power group.
    }
    \label{ff:2tx}
\end{figure}

Fig.~\ref{ff:2tx} and Table~\ref{tab:2} highlight the consistently superior performance of PLAN-Net over all other models and benchmarks in the grid Wi-Fi network.
When $P_\text{tx}$ is increased (Fig.~\ref{ff:2tx}), the signal can reach farther nodes, reducing the need for transmission relays and shortening transmission paths, ultimately reducing \revis{the mean and variance of delays}.
It is also worth discussing the increased performance advantage that PLAN-Net has gained over ns-3 compared to the wired setting, as is evident in all KPIs (Table~\ref{tab:2}). 
In a network with a regular grid topology, it is likely that multiple shortest paths exist between a source node and a destination node. 
The ns-3 routing scheme, which uses a shortest path algorithm, may not always choose the same path each time due to this indeterminacy. 
Hence, the ns-3 benchmarks might be using a routing instance different than the one used to generate the ground-truth KPIs.
This naturally leads to increased testing errors in the ns-3 benchmarks.
However, learning-based models, such as PLAN-Net, do not suffer from this issue since they take the specific routing instance as an explicit input. 
This means that they are not confused by the regularity of the grid shape and the existence of multiple possible shortest paths.

\begin{table}[t]
    \caption{
    Performance of candidate methods for multiple KPIs measured by NMAE on the test set.
    The transmit power for both the training and test samples is fixed at $16\text{ dBm}$.
    }\label{tab:2}
    \centering
    \small
    \resizebox{\linewidth}{!}{%
        \begin{tabular}{|l|cc|cc|cc|cc|}
        \hline
             \multirow{2}{*}{\diagbox[width=2cm]{Method}{Metric}} &  \multicolumn{2}{c}{Delay} &  \multicolumn{2}{c}{Jitter} &  \multicolumn{2}{c}{Throughput} &  \multicolumn{2}{c|}{Drops} \\
        \cline{2-9}
             & Mean  & SD & Mean  & SD & Mean  & SD & Mean  & SD \\
        \hline
PLAN-Net & \textbf{0.23}  & \textbf{0.28} & \textbf{0.35}  & \textbf{0.50} & \textbf{0.19}  & \textbf{0.18} & \textbf{0.24}  & \textbf{0.27} \\
RouteNet & 0.31  & 0.32 & 0.42  & 0.55 & 0.23  & 0.22 & 0.32  & 0.35 \\
GNN & 0.25  & 0.32 & 0.41  & 0.54 & 0.27  & 0.22 & 0.26  & 0.28 \\
ns-3 & 0.32  & 0.39 & 0.46  & 0.88 & 0.25  & 0.26 & 0.33  & 0.41 \\
ns-3$^{+}$ & 0.27  & 0.34 & 0.40  & 0.71 & 0.22  & 0.22 & 0.28  & 0.34 \\
ns-3$^{+\!+}$ & 0.26  & 0.32 & 0.38  & 0.65 & 0.20  & 0.21 & 0.27  & 0.32 \\
        \hline
        \end{tabular}
    }
    \vspace{-0.5em}
\end{table}

\subsection{Wireless network with topological perturbations}\label{ss:wifi-pert}

\begin{figure}[t]
    \centering
    \begin{subfigure}[t]{\linewidth}
        \centering
        \includegraphics[width=\linewidth, trim=0 8cm .3cm 0, clip]{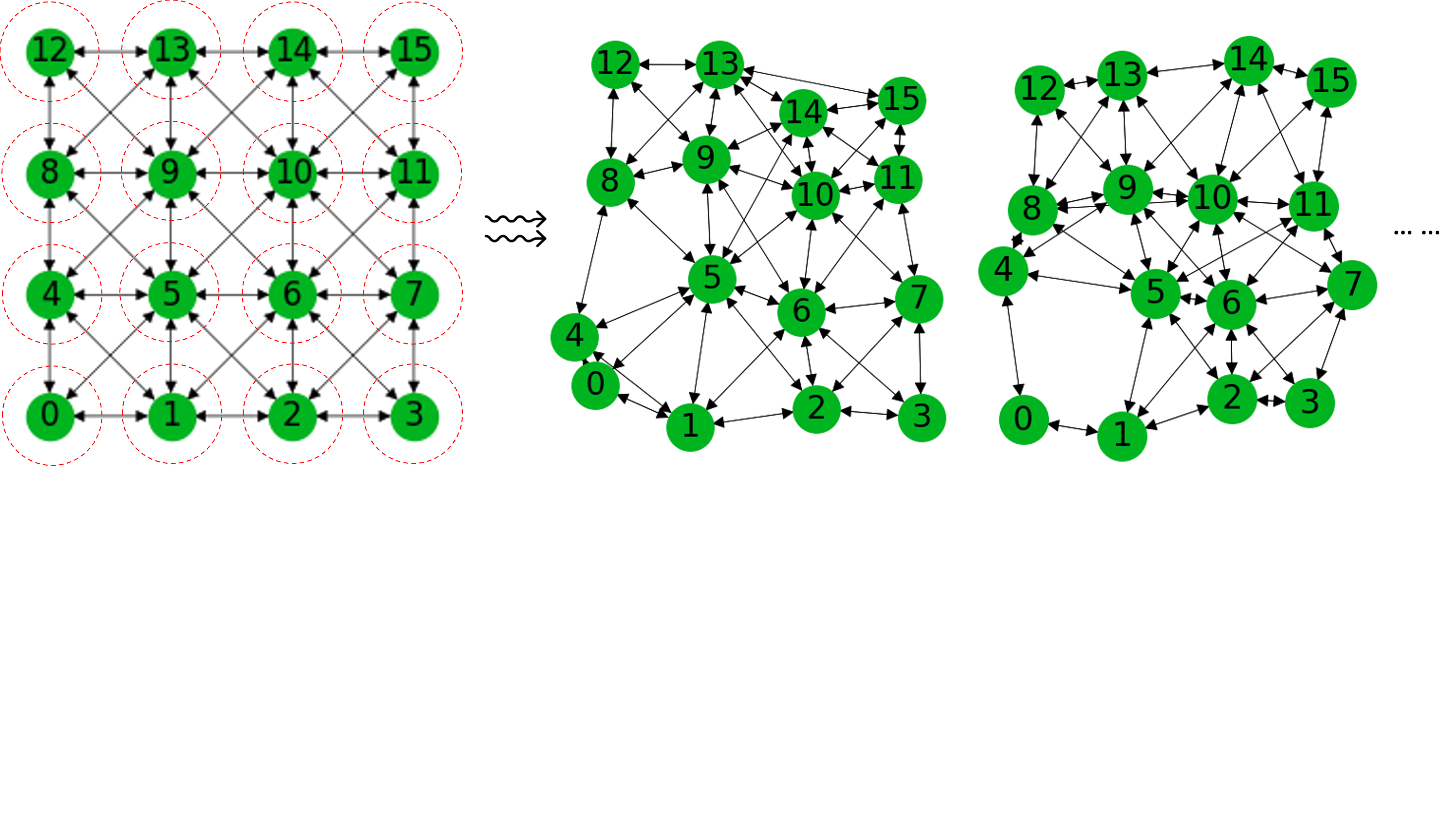}
    \end{subfigure}%
    \caption{
    Perturbed grid Wi-Fi network topologies. 
    }
    \label{ff:wifi-topo-per-illu}
\end{figure}

In our final experiment, we aim to assess the robustness and generalization capabilities of PLAN-Net by introducing position perturbations to the regular grid topology. 
By randomly repositioning nodes within a 10-meter radius centered at their original locations (Fig.~\ref{ff:wifi-topo-per-illu}), we simulate a more realistic wireless network environment where node positions may change over time. 
As a result, some links between nodes may be lost due to increased distance beyond the communication range, while new links may be established as some nodes move closer together. 
These changes not only make learning more challenging across different topologies but also result in less  indeterminacy in routing tables, which allows ns-3 to regain some advantages compared to the regular grid.

Table~\ref{tab:3} shows that the delay, jitter, and throughput prediction performance of PLAN-Net is comparable to ns-3$^+$ when applied to the perturbed grid topology.
Even for the least-performing prediction on drops, PLAN-Net still matches the single-run ns-3 performance.
Interestingly, by incorporating node embeddings, PLAN-Net achieves a remarkable improvement in performance with respect to RouteNet, with a less than $9\%$ growth in trainable parameters and negligible inference time increase on the scale of milliseconds.

\begin{table}[t]
    \caption{
    Performance of candidate methods in predicting multiple KPIs measured by NMAE on the test set.
    The training and test samples have different topologies due to positional perturbations.
    }\label{tab:3}
    \centering
    \small
    \resizebox{\linewidth}{!}{%
        \begin{tabular}{|l|cc|cc|cc|cc|}
        \hline
             \multirow{2}{*}{\diagbox[width=2cm]{Method}{Metric}} &  \multicolumn{2}{c}{Delay} &  \multicolumn{2}{c}{Jitter} &  \multicolumn{2}{c}{Throughput} &  \multicolumn{2}{c|}{Drops} \\
        \cline{2-9}
             & Mean  & SD & Mean  & SD & Mean  & SD & Mean  & SD \\
        \hline
PLAN-Net & 0.29  & 0.31 & 0.46  & 0.71 & 0.23  & 0.21 & 0.29  & 0.31 \\
RouteNet & 0.34  & 0.34 & 0.48  & 0.74 & 0.25  & 0.24 & 0.33  & 0.37 \\
GNN & 0.32  & 0.35 & 0.54  & 0.77 & 0.29  & 0.24 & 0.30  & 0.32 \\
ns-3 & 0.33  & 0.44 & 0.51  & 1.09 & 0.25  & 0.27 & 0.30  & 0.38 \\
ns-3$^{+}$ & 0.29  & 0.37 & 0.45  & 0.86 & 0.22  & 0.23 & 0.26  & 0.32 \\
ns-3$^{+\!+}$ & \textbf{0.27}  & \textbf{0.35} & \textbf{0.43}  & \textbf{0.80} & \textbf{0.21}  & \textbf{0.22} & \textbf{0.25}  & \textbf{0.30} \\
        \hline
        \end{tabular}
    }
    \vspace{-0.5em}
\end{table}

\vspace*{1mm}
Overall, these experiments allow us to evaluate how well PLAN-Net can adapt to different underlying topologies and provides insights into the performance trade-offs between learning-based models and simulation-based methods.
The obtained results indicate that PLAN-Net can achieve a performance similar to the one obtained by averaging several runs of ns-3.
However, it should be noted that PLAN-Net's KPI estimates require only a forward pass of our architecture (10s - 100s of milliseconds) whereas several runs of ns-3 require 100s of seconds. 
This entails \emph{a gain of 3 to 4 orders of magnitude in computation time}.

\section{Conclusions and Future Work}\label{s:end}

PLAN-Net extended RouteNet's capabilities achieving better generalization results and higher accuracy. 
The introduction of node embeddings helped to differentiate parallel and star topologies for path-link predictions and enabled the implicit consideration of interference effects. 
The most promising area of application is network configuration optimization, given the model's ability to generalize to unseen topologies and make quick predictions for them. 
In the future, we will extend the initial embedding features and introduce more topologies, such as hierarchical wireless networks. 
Additionally, we plan to explore the effectiveness of PLAN-Net in diverse scenarios, including mobile relay healing of wireless networks and KPI-informed control of mobility.

% \vfill\pagebreak
\bibliographystyle{IEEEbib}
{
    \footnotesize
    \bibliography{IEEEabrv,references}
}

\end{document}